# Near-real-time monitoring of global ocean carbon sink

Piyu Ke[1,2,3,*], Xiaofan Gui[3], Wei Cao[3], Dezhi Wang[4], Ce Hou[5,6], Lixing Wang[1], Xuanren Song[1], Yun Li[7], Biqing Zhu[8], Jiang Bian[3], Stephen Sitch[2], Philippe Ciais[9,10], Pierre Friedlingstein[2], Zhu Liu[1,11,*]

[1] *Department of Earth System Science, Tsinghua University, Beijing, China*
[2] *Department of Mathematics and Statistics, Faculty of Environment, Science and Economy, University of Exeter, Exeter, UK*
[3] *Microsoft research*
[4] *School of Mathematics and Statistics, Lanzhou University, Lanzhou, China*
[5] *Department of Civil and Environmental Engineering, The Hong Kong University of Science and Technology, Hong Kong SAR, China*
[6] *Institute of Remote Sensing and Geographical Information System, School of Earth and Space Sciences, Peking University, Beijing, China*
[7] *Department of architecture, faculty of engineering science, KU Leuven, Leuven, Belgium*
[8] *Integrated Assessment and Climate Change Research Group and Exploratory Modeling of Human-natural Systems Research Group, International Institute for Applied Systems Analysis (IIASA), 2361 Laxenburg, Austria*
[9] *Laboratoire des Sciences du Climat et de l'Environnement LSCE, Orme de Merisiers 91191 Gif-sur-Yvette, France*
[10] *Climate and Atmosphere Research Center (CARE-C) The Cyprus Institute 20 Konstantinou Kavafi Street, 2121, Nicosia, Cyprus*
[11] *Institute of Climate and Carbon Neutrality, Department of Geography, The University of Hong Kong, Hong Kong SAR, China*

These authors contributed equally: Piyu Ke, Xiaofan Gui.

*Corresponding authors:

Piyu Ke (kpy20@mails.tsinghua.edu.cn), Zhu Liu (zhuliu@tsinghua.edu.cn)

## Abstract

The ocean plays a critical role in modulating climate change by absorbing atmospheric CO2. Timely and geographically detailed estimates of the global ocean-atmosphere CO2 flux provide an important constraint on the global carbon budget, offering insights into temporal changes and regional variations in the global carbon cycle. However, previous estimates of this flux have a 1 year delay and cannot monitor the very recent changes in the global ocean carbon sink. Here we present a near-real-time, monthly grid-based dataset of global surface ocean fugacity of CO2 and ocean-atmosphere CO2 flux data from January 2022 to July 2023, which is called Carbon Monitor Ocean (CMO-NRT). The data have been derived by updating the estimates from 10 Global Ocean Biogeochemical Models and 8 data products in the Global Carbon Budget 2022 to a near-real-time framework. This is achieved by employing Convolutional Neural Networks and semi-supervised learning methods to learn the non-linear relationship between the estimates from models or products and the observed predictors. The goal of this dataset is to offer a more immediate, precise, and comprehensive understanding of the global ocean-atmosphere CO2 flux. This advancement enhances the capacity of scientists and policymakers to monitor and respond effectively to alterations in the ocean's CO2 absorption, thereby contributing significantly to climate change management.

## Background & Summary

The ocean plays a crucial role in absorbing anthropogenic heat and carbon. The annual Global Carbon Budget report includes estimates of global ocean carbon sink, that lag behind actual conditions by 1 year[1]. As global efforts to mitigate climate change intensify, and in support of the global stocktake process crucial for progressing towards the objectives of the Paris Climate Change Agreement, there is a rising need for more accurate data on global ocean carbon sinks with a lower latency. Numerous researchers and institutions have developed databases for surface ocean fugacity of CO2 (fCO2), surface ocean partial pressure of CO2 (pCO2) and air-sea CO2 flux. The latest update of the community-led Surface Ocean CO2 Atlas (www.socat.info), version 2023, includes 35.6 million quality-controlled, in situ surface ocean fCO2 measurements, collected between 1957 and 2022[2]. Observation-based products employ the sparse pCO2 observations from SOCAT, utilizing multivariate linear regression or machine learning algorithms to connect these data with comprehensive observations of related variables[3-10]. This methodology enables the estimation of pCO2 at every spatial and temporal point. Global Ocean Biogeochemical Models (GOBMs) comprehensively simulate the dynamics of the ocean's carbonate system, encompassing its physics, biology, and chemistry[11-21]. These models are grounded in equations that embody the physical and biogeochemical processes of the ocean. By being driven with observed wind patterns and surface energy fluxes, they can accurately estimate the ocean's physical and biogeochemical states over recent decades. The outputs of these models encompass a wide range of variables, such as surface ocean pCO2 and the flux of CO2 between the air and sea. However, the global ocean carbon sink estimates provided by these methodologies are only up-to-date as of December 2021, causing an approximate two-year delay in current data availability. This delay stems primarily from the significant computational demands of GOBMs, which naturally result in some latency. Additionally, observation-based products depend on the SOCAT for generating their results. Given that SOCAT data itself typically experiences a 1 year lag, this further contributes to delays in updating the machine learning model data.

Here, we introduce Carbon Monitor Ocean (CMO-NRT), a near-real-time dataset providing monthly gridded global surface ocean fCO2 and ocean-atmosphere CO2 flux data, spanning January 2022 to July 2023 (Fig. 1). This paper presents th methodology and findings, outlining data sources and approaches in the Methods section and providing an evaluation of CMO-NRT in the Technical Validation section. The dataset is publicly accessible on Figshare[22] and our website at https://carbonsink.microsoft.com/.

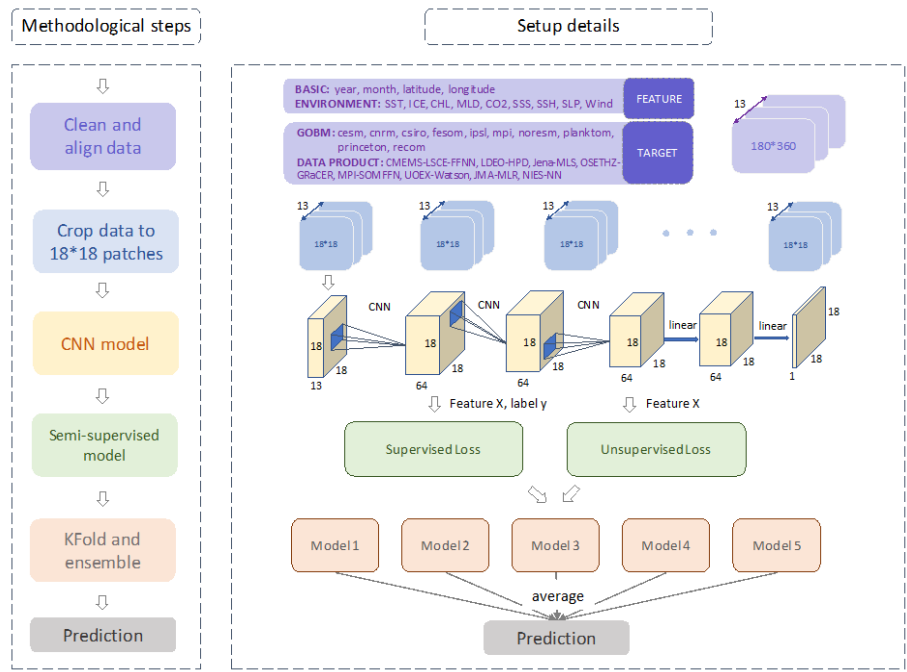

**Figure 1**. Schematic overview of the methodology and data sources for Carbon Monitor Ocean (CMO-NRT).

# Methods

Carbon Monitor Ocean (CMO-NRT) utilizes advanced deep learning methods to build models that connects the output of GOBMs or ocean data products with relevant environmental variable observations. We then update the estimates from 10 GOBMs and 8 data products in the Global Carbon Budget 2022, providing each model and product with monthly gridded estimations from January 2022 to July 2023. A schematic of CMO-NRT is shown in Figure 2. Below we describe the calculation process in detail.

**Data sources and pre-processing**

Global ocean biogeochemical models and observation-based data products

GOBMs and data products serve as benchmarks for ocean fCO2 estimations. We utilize monthly data from 10 GOBMs and 8 data products that contributed to the Global Carbon Budget 2022[1,23], with data up to the end of 2021 (Table 1). Meteorological reanalysis and atmospheric CO2 levels are applied to drive each model. Each GOBM employs a set of interconnected differential equations to model the physical, chemical, and biological factors affecting surface ocean pCO2. While each data product relies on multivariate linear regression or machine learning techniques to establish a model that correlates SOCAT observational data with relevant variable observations. The fCO2 output from each GOBM or data product is provided at a 1° × 1° monthly resolution.

Observed predictors

Our predictive variables include a range of biological, chemical, and physical factors that are typically linked to fluctuations in fCO2. These variables are sea surface temperature (SST), sea ice fraction (ICE), sea surface salinity (SSS), atmospheric CO2 mole fraction (xCO2), mixed-layer depth (MLD), sea surface height (SSH), chlorophyll a (chl a), sea level pressure (SLP), and wind speed. The data sources for these variables are detailed in Table 2. All data are bilinearly interpolated from their original grid to a 1° × 1° monthly resolution, aligning with our fCO2 targets.

Given that the xCO2 data is only available up to the end of 2022, and to meet the requirement for near-real-time data, we gather global average marine boundary layer surface monthly mean atmospheric CO2 data updated to July 2023, and employ a light gradient boosting machine (LightGBM)[24] model to establish a relationship between the year, month, latitude, longitude, mean atmospheric CO2 data and xCO2. We used data from 1979-2021, divided into training and validation datasets at an 8:2 ratio. We implemented early stopping with LightGBM, and then tested on 2022 data, obtaining a test RMSE of 1.74, reflecting roughly a 0.5% prediction error. This approach allows us to extend the xCO2 data to near-real-time.

| Type | Datasets | Data information |
|---|---|---|
| Global ocean biogeochemistry models | NEMO-PlankTOM12[11] | Surface ocean fugacity of $CO_2$, 1959/01-2021/12, monthly, 180*360 |
| | MICOM-HAMOCC (NorESM-OCv1.2)[12] | |
| | MPIOM-HAMOCC6[13] | |
| | NEMO3.6-PISCESv2-gas (CNRM)[14] | |
| | FESOM-2.1-REcoM2[15] | |
| | MOM6-COBALT (Princeton)[16] | |
| | CESM-ETHZ[17] | |
| | NEMO-PISCES (IPSL)[18] | |
| | MRI-ESM2-1[19,20] | |
| | CESM2[21] | |
| Ocean data products | MPI-SOMFFN[3] | |
| | Jena-MLS[4] | |
| | CMEMS-LSCE-FFNNv2[5] | |
| | LDEO-HPD[6] | |
| | UOEx-Watson[7] | |

| | NIES-NN[8] | |
| | JMA-MLR[9] | |
| | OS-ETHZ-GRaCER[10] | |

**Table 1**. Global ocean biogeochemical models and ocean data products used in the Global Carbon Budget 2022[1,23].

| Variable | Abbreviation | Data Product | Resolution |
|---|---|---|---|
| Sea Surface Temperature | SST | NOAA: OISST[25] | 1981/9-2023/07, daily, 720*1440 (latitude*longitude) |
| Sea Ice Fraction | ICE | | |
| Sea Surface Salinity | SSS | Met Office: EN4[26] | 1959/01-2023/07, monthly, 173*360 |
| Atmospheric $CO_2$ mixing ratio | $xCO_2$ | NOAA: Greenhouse Gas Marine Boundary Layer Reference[27] | 1979/01-2022/12, weekly, 180*1 |
| Mixed Layer Depth | MLD | ECMWF: ORAS5[28] | 1959/01-2023/08, monthly, 1021*1442 |
| Sea Surface Height | SSH | | |
| Chlorophyll-a | Chl a | ESA: GlobColour[29] | 1997/09-2023/09, monthly, 180*360 |
| Sea Level Pressure | SLP | ECMWF: ERA5[30] | 1959/01-2023/08, monthly, 1021*1442 |
| Wind Speed | Wind | | |
| Year, month, longitude and latitude | | | |

**Table 2**. Sources of input data sets.

## Deep learning method

Our study utilized a deep learning-based technique tailored for the near-real-time estimation of oceanic carbon monthly gridded fluxes (Fig. 2). The approach integrates year, month, latitude, longitude and nine environmental factor data as inputs, with a GOBM model or an ocean data product serving as the prediction targets. Each set of data was processed into a 180x360 grid format. For computational efficiency, we dissected all environmental factors into 18x18 patches. Data points featuring a specific label value were designated as $D_l$, while those with null labels were termed $D_u$. Initially, we trained the model on all $D_l$ data points, calculating the Root Mean Square Error (RMSE) to serve as the supervised loss $L_s$. For the combined $D_l$ and $D_u$ data set, we partitioned it into two subsets: one with 10% of features randomly removed for weak augmentation, and another with 30% of features randomly eliminated for strong enhancement. The model generated predictions for both subsets separately, and the RMSE between the two predicted data sets was computed to serve as the unsupervised loss $L_u$. The model was subsequently updated through backward propagation, utilizing the sum of $wL_u + L_s$ as the model's loss. This method has been shown to enhance the model's stability. The structural design of the model, as outlined in the bottom right corner of Fig. 1, incorporates a stacking of multilayer Convolutional Neural Network[31]

(CNN) models and linear models[32]. The input layer aligns with the patch format, having a dimension of 18x18x9, with the width and height being consistent across all CNN and linear layers. The hidden layers within the CNN model are configured with dimensions set at 9, 64, and 64. The linear model layers are structured with dimensions of 64, 64, and 1. The model culminates in an output layer with a dimension of 1.

Convolutional Neural Network (CNN), inspired by the human brain's functionality, particularly visual perception, are deep learning models engineered to analyze grid-like structured data, such as images or videos. This network comprises several layers: Convolutional Layer, Rectified Linear Unit (ReLU) Layer, and Fully Connected Layer. In the context of our study, we have configured the hidden dimension to 64 with a kernel size of 3 in the convolutional layer. In the ReLU layer, we have incorporated the Gaussian Error Linear Unit (GELU)[33] activation function, followed by a dropout strategy with a probability (p) of 0.25 to prevent overfitting.

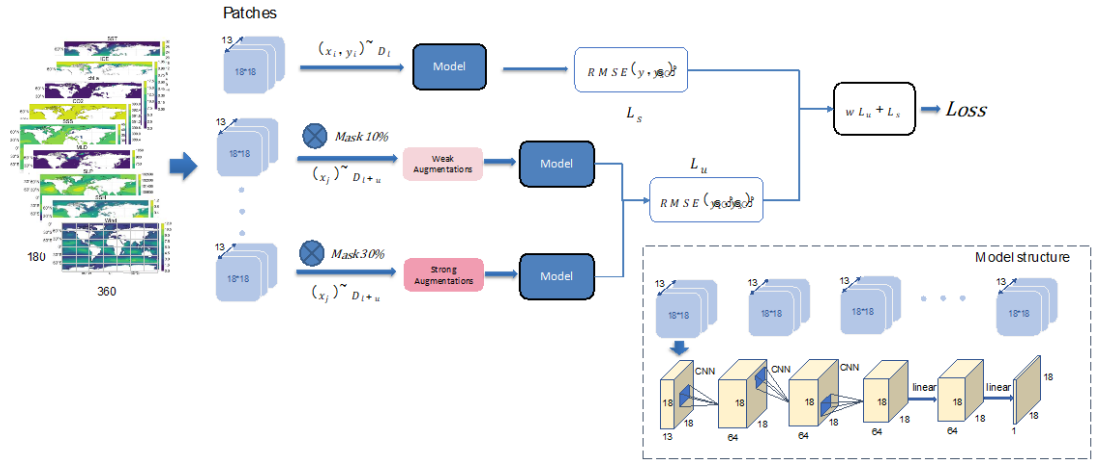

**Figure 2.** Schematic diagram of our methodology.

**Calculations of pCO₂ and air-sea CO₂ flux**

The pCO2 is calculated by the following equation:

$$pCO_2 = fCO_2 \times exp(P_{atm}^{surf} \times \frac{B + 2\delta}{R \times T})^{-1}$$

where $P_{atm}^{surf}$ is the atmospheric surface pressure from ECMWF Reanalysis version 5[30], $T$ is the sea surface temperature (SST) in Kelvin from National Oceanic and Atmospheric Administration (NOAA) optimally interpolated SST (OISST)[25], $B$ and $\delta$ are virial coefficients from Weiss[34], $R$ is the gas constant[35].

The air–sea CO₂ flux is calculated here by the standard bulk equation:

$$F_{CO_2} = k_w \times S_{CO_2} \times (1 - f_{ice}) \times (pCO_2^{atm-moist} - pCO_2^{ocean})$$

which parameterizes the air-sea CO₂ flux ($F_{CO_2}$) as a function of the gas transfer velocity ($k_w$), CO₂ solubility ($S_{CO_2}$), ice fraction ($f_{ice}$), and partial pressure of CO₂ in moist air ($pCO_2^{atm-moist}$) and surface ocean ($pCO_2^{ocean}$). Solubility is calculated following Weiss[34] and partial pressure of moist air ($pCO_2^{atm-moist}$) is calculated following Equation,

$$pCO_2^{atm-moist} = xCO_2 \times (P_{atm} - pH_2O)$$

where $xCO_2$ is the dry air mixing ratio of atmospheric $CO_2$ from NOAA Greenhouse Gas Marine Boundary Layer Reference[27], $P_{atm}$ is the total atmospheric pressure from ECMWF Reanalysis version 5[30], and $pH_2O$ is the saturation vapor pressure[35]. We use the Wanninkhof[36] formulation for the gas transfer velocity:

$$k_w = k_{w,scaled} \times u^2 \times (\frac{S_c}{660})^{-0.5}$$

which parameterizes $k_w$ as a function of wind speed squared ($u^2$) and the Schmidt number ($S_c$). $k_w$ is scaled by a factor of $k_{w,scaled}$ for each wind product to match the invasion of bomb $^{14}C$ (343 ± 40 × 1026 atoms $^{14}C$ as of 1994)[37,38]. The wind product is from ECMWF Reanalysis version 5[30].

## Data Records

The CMO-NRT dataset, formatted in Network Common Data Form (NetCDF), provides monthly global oceanic surface fCO2 and air-sea CO2 flux data. This dataset spans from January 2022 through July 2023 and is structured into a 1° x 1° grid. It includes three dimensions and two variables.

The dimensions are as follows:

Time: Monthly data, from January 2022 to July 2023.

Latitude (lat): Ranges from -90° to 90° North.

Longitude (lon): Spans from -180° to 180° East.

Product: Involves 10 GOBMs and 8 data products.

The two variables covered are:

sfco2: This represents the surface ocean fCO2, quantified in units of uatm. It is measured across 18 products, 19 time points, and a grid of 180 latitudes by 360 longitudes.

fgco2: This indicates the flux density of total air-sea CO2 exchange, expressed in gC/m²/day with a positive value indicating an upward direction. Similar to sfco2, it is measured across the same dimensions.

As of the time of this writing, the dataset includes data up to July 2023. The complete dataset is accessible for download on Figshare[22]. For the latest updates and additional information, visit our website at https://carbonsink.microsoft.com/.

Figure 3 displays the monthly oceanic fCO2 and air-sea CO2 flux from 1990 through July 2023, alongside their corresponding gridded data for the period of August 2022 to July 2023 as estimated by CMO-NRT.

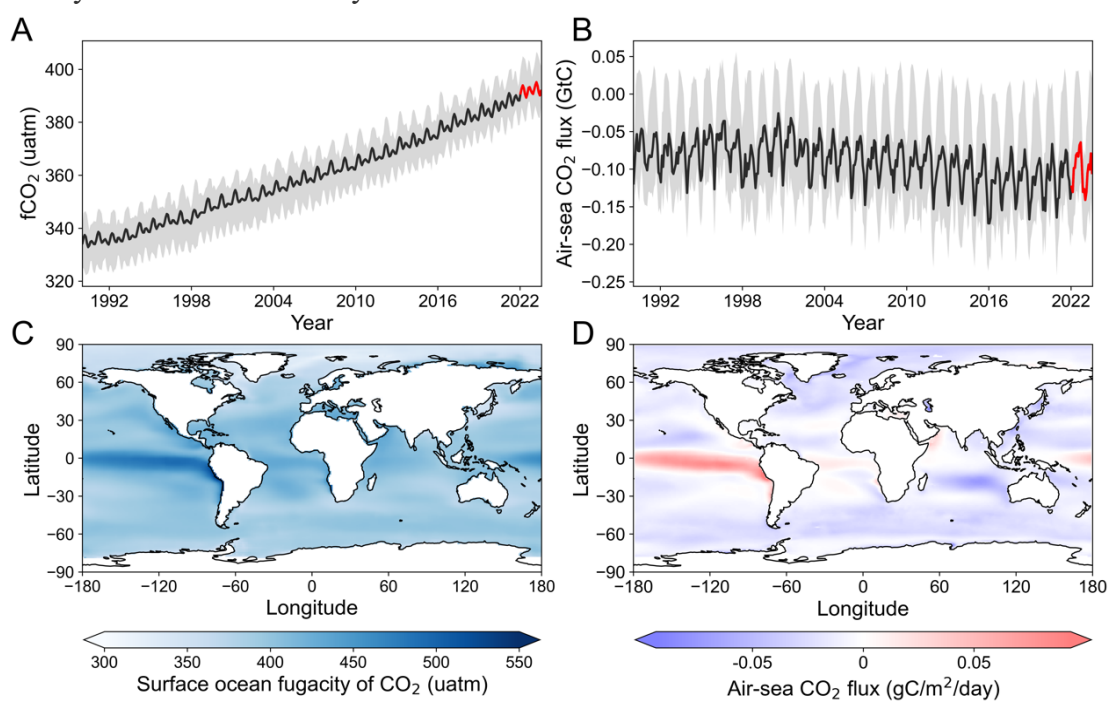

**Figure 3**. Monthly fCO2 (A) and air-sea CO2 flux (positive upward) (B) average of 10 GOBMs and 8 data products over 1990-2021 from Global Carbon Budget 2022 (black lines) and 2022-July 2023 (near-real-time predictions, red lines) from Carbon Monitor Ocean (CMO-NRT). Grey shaded areas represent the range of estimations from 10 GOBMs and 8 data products. Mean fCO2 (C) and air-sea CO2 flux (D) over August 2022-July 2023 estimated from CMO-NRT.

## Technical Validation

To evaluate the near-real-time predictions of CMO-NRT, we excluded the most recent two years' data of the 10 GOBMs and 8 data products. Using the approach outlined in this study, we trained models to learn the non-linear correlations between the fCO2 values from GOBMs or data products and observed predictors during 2000-2019. We then produced monthly fCO2 predictions for 2020-2021. Then we can compare these predictions with the original outputs for the same period. Our comparative analysis was conducted from three distinct perspectives: correlation, the global quantity and spatial distribution.

**Correlation**

Figures 5 and 6 display the correlation between CMO-NRT predictions and original outputs for each of the 10 GOBMs and the 8 data products during 2020-2021. Overall, the predictions for both GOBMs and data products demonstrated a strong performance, with most having an $R^2$ value above 0.9. Scatter plot comparisons reveal that, in data products, there are occasional points where predictions significantly deviate from the fit line. In contrast, for GOBMs, most points cluster close to the fit line, indicating more stable performance of the model with GOBMs.

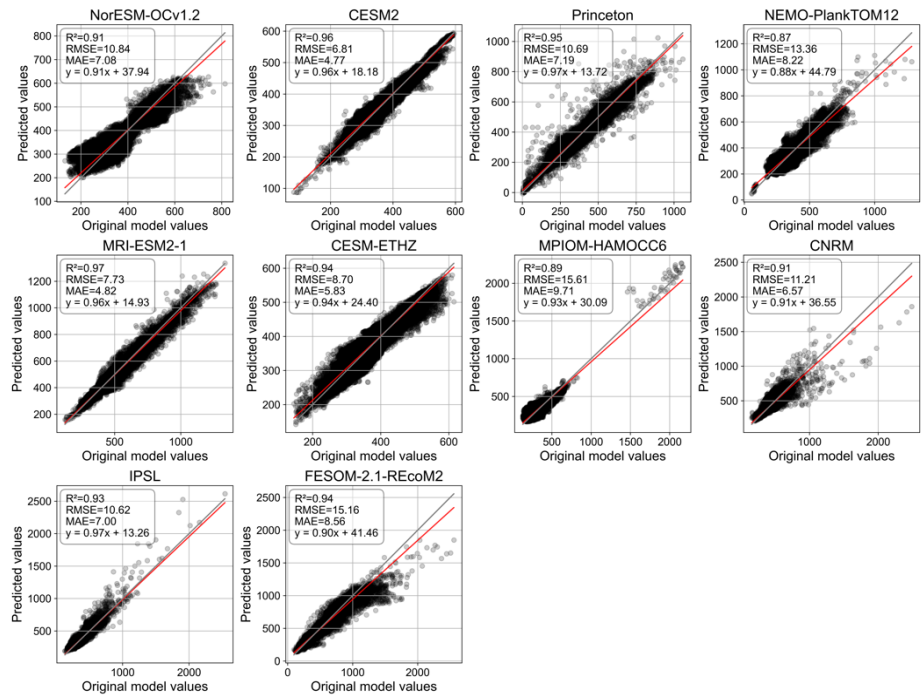

**Figure 5**. Correlation between CMO-NRT predictions and original outputs for each of the 10 GOBMs during 2020-2021. The red lines show the best-fit line from total least-squares regression. The grey lines represent the 1:1 regression.

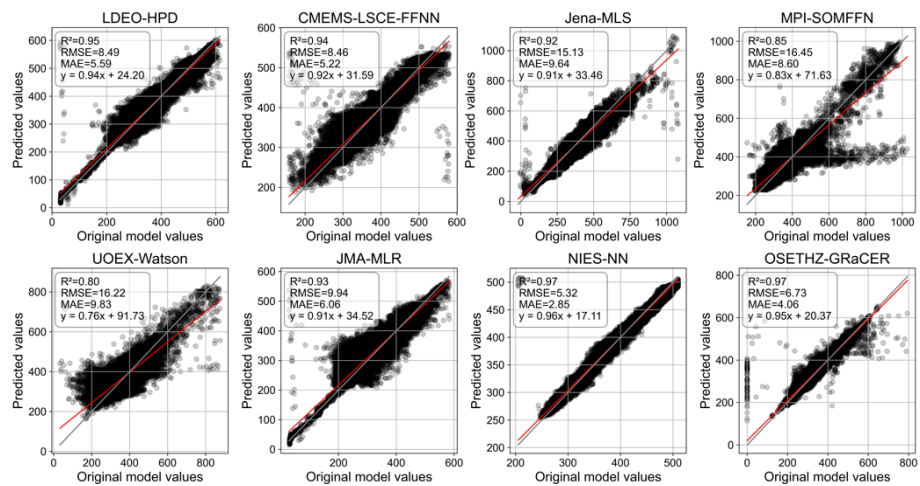

**Figure 6**. Correlation between CMO-NRT predictions and original outputs for each of the 8 data products during 2020-2021. The red lines show the best-fit line from total least-squares regression. The grey lines represent the 1:1 regression.

## Global quantity

We further analyzed the monthly variations of global fCO2, comparing CMO-NRT predictions with original outputs during 2020-2021 as shown in Figures 7 and 8. The results exhibit a high degree of agreement in terms of seasonality across both GOBMs and data products. Generally, our estimated values were slightly higher than the original values, with most differences being less than 3 µatm. In terms of global aggregate comparisons, GOBMs demonstrated superior performance, with most $R^2$ values exceeding 0.85.

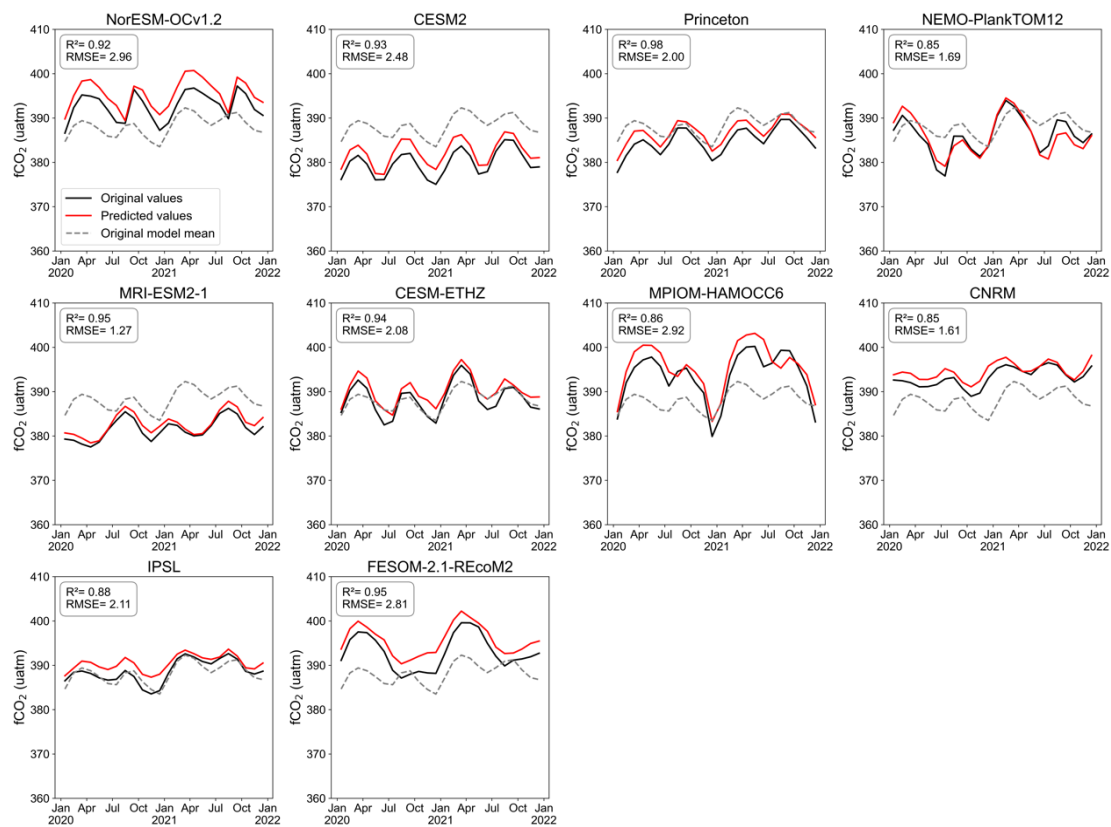

**Figure 7**. Comparison of global average monthly fCO2 between CMO-NRT predictions and original outputs for each of the 10 GOBMs during 2020-2021. The red lines are the predicted results from CMO-NRT. The black lines are the original results from GOBMs. The dashed grey lines are the average results of 10 GOBMs.

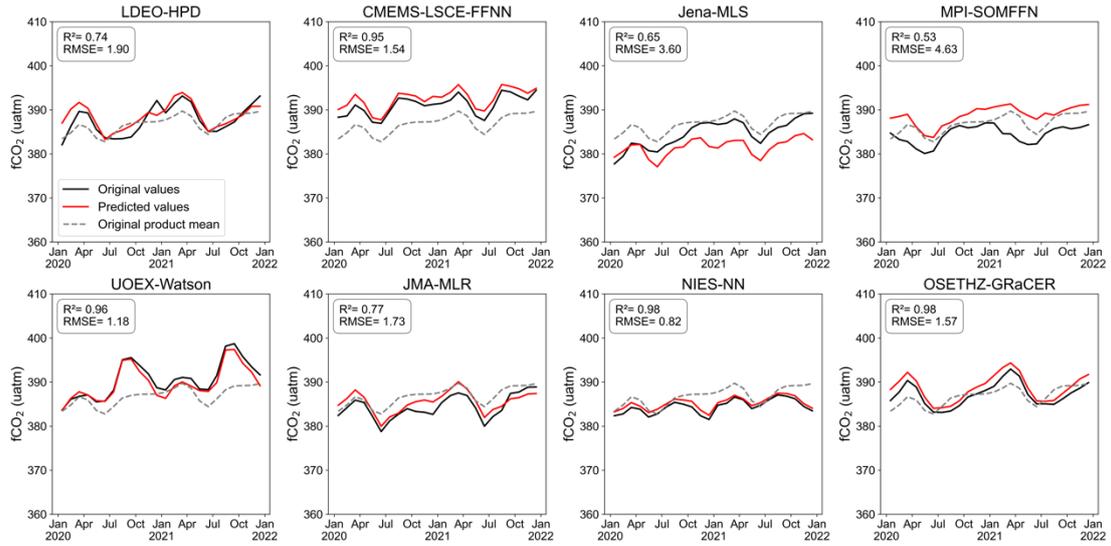

**Figure 8**. Comparison of global average monthly fCO2 between CMO-NRT predictions and original outputs for each of the 8 data products during 2020-2021. The red lines are the predicted results from CMO-NRT. The black lines are the original results from data products. The dashed grey lines are the average results of 8 data products.

## Spatial distribution

We evaluate the spatial patterns of mean fCO2 of the CMO-NRT prediction against the original output for each of the 10 GOBMs and 8 data products during 2020-2021 (Fig. 9 and Fig. 10). The spatial variations in fCO2 were largely consistent between the CMO-NRT predictions and the original results, with most discrepancies being under 20 μatm. The predicted values also align well with the original outputs in both trend and magnitude across latitudes. Notably larger errors were observed in the equatorial Pacific, including adjacent coastal areas near Peru and Chile to the west and northeast regions near Indonesia and Papua New Guinea, as well as the Arctic Ocean. These errors are primarily due to these regions typically exhibiting extreme maximum and minimum fCO2 values. Our model tends to perform less accurately in predicting extremes compared to values closer to the mean, resulting in slightly poorer performance in these areas. Additionally, the MPI-SOMFFN shows lower predictive accuracy in the Arctic region, primarily due to the high frequency of missing historical data in this area.

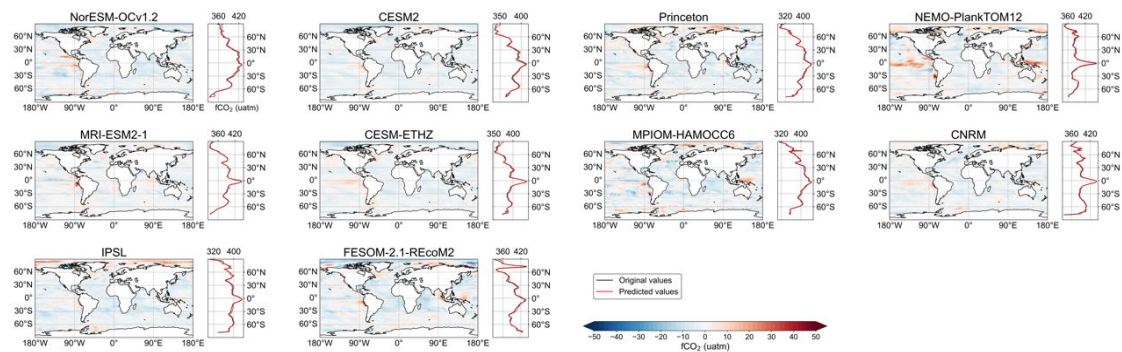

**Figure 9**. Spatial differences between CMO-NRT predictions and original outputs for each of the 10 GOBMs during 2020-2021. The line graphs represent the values of predictions and original outputs across latitudes.

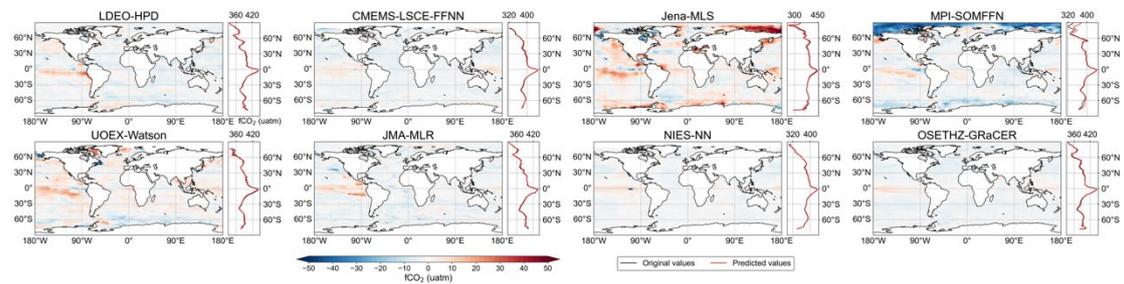

**Figure 10**. Spatial differences between CMO-NRT predictions and original outputs for each of the 8 data products during 2020-2021. The line graphs represent the values of predictions and original outputs across latitudes.

We also calculated the monthly mean fCO2 values by averaging them separately across the 10 GOBMs and 8 data products during 2020-2021. The spatial variation in fCO2 showed high consistency between the CMO-NRT predictions and the original results, with discrepancies generally under 10 μatm for GOBMs and data products. The predicted values align well with the original outputs in both trend and magnitude across latitudes. Examining different months, we observed that in the summer months, the original values from GOBMs in the Arctic region were significantly higher than our predictions. In contrast, the performance of data products was more uniform across different months. Typically, the original values from data products were lower than the predictions in the Arctic Ocean and higher in the equatorial Pacific region.

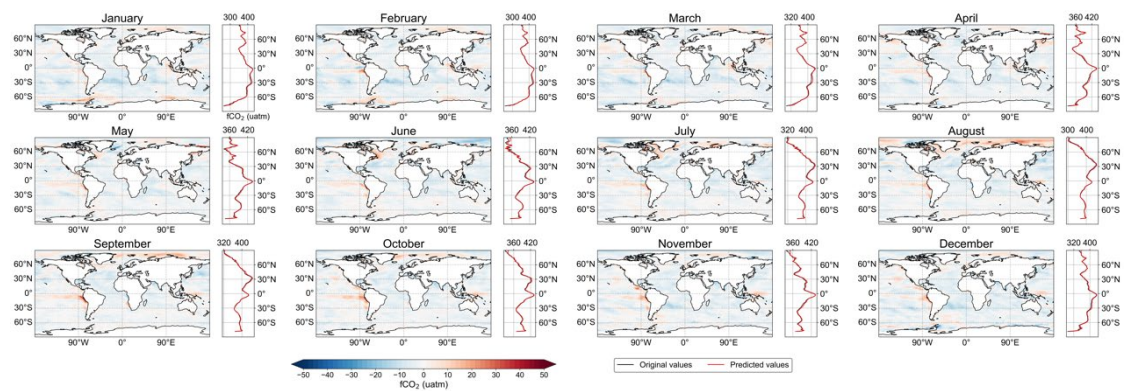

**Figure 11**. Spatial differences between CMO-NRT predictions and original outputs of the average of 10 GOBMs across 12 months during 2020-2021. The line graphs represent the values of predictions and original outputs across latitudes.

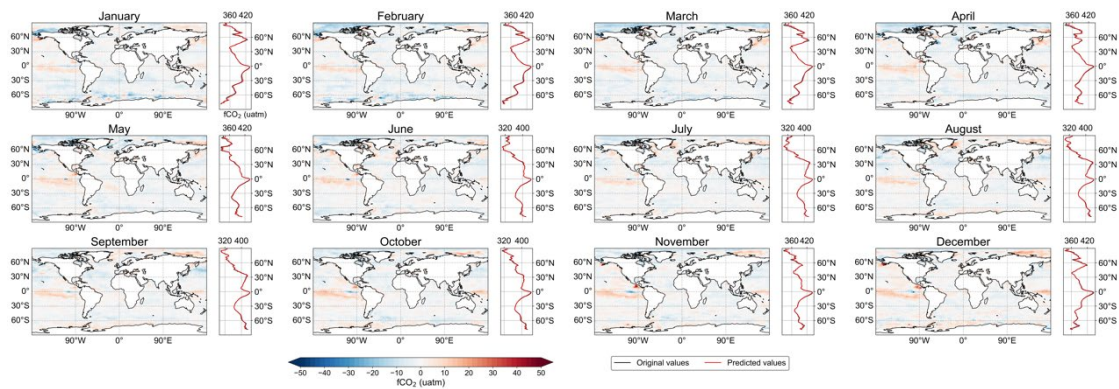

**Figure 12**. Spatial differences between CMO-NRT predictions and original outputs of the average of 8 data products across 12 months during 2020-2021. The line graphs represent the values of predictions and original outputs across latitudes.

## Usage Notes

The generated datasets are available from https://doi.org/10.6084/m9.figshare.24658494.v1. Latest updates and related information are available for view and download on our website https://carbonsink.microsoft.com/.

## Code availability

Python code for producing the model and datasets is provided at https://github.com/kepiyu/CMO-NRT.

## Acknowledgements

We acknowledge the National Natural Science Foundation of China (grant 71874097, 41921005, 72140002 and 72140002), Beijing Natural Science Foundation (JQ19032), and the Qiu Shi Science & Technologies Foundation. We thank the Global Carbon Project and the model developers for use of the Global Ocean Biochemical Models (GOBMs) and data products.

## Author contributions

P.K., X.G. and Z.L. designed the research. P.K. X.G and W.C. designed the methods and conducted the data processing. P.K. and X.G. wrote the manuscript, and all authors contributed to data collection, discussion and analysis.

## Competing interests

The authors declare no competing interests.